\documentclass[letterpaper,12pt]{article}

\usepackage{graphicx}
\usepackage{subcaption}
\usepackage{amssymb}
\usepackage{float}
\usepackage{longtable}
\usepackage{amsmath}
\usepackage{array}
\usepackage{tabularx}
\usepackage{enumitem}
\usepackage{color}
\usepackage{amsmath,textcomp,amssymb,geometry,graphicx,enumerate}
\usepackage{algorithm} 
\usepackage[noend]{algpseudocode} 
\usepackage{mathtools}
\newcounter{descriptcount}

\usepackage{pbox}
\usepackage{xcolor}
\usepackage{titlesec}
\graphicspath {{figures/}}
\usepackage{moreverb}

\usepackage{mathptmx}

\usepackage{graphics}

\usepackage{natbib}

\begin{document}


\title{Propensity score prediction for electronic healthcare databases using Super Learner and High-dimensional Propensity Score Methods}
\author{Cheng Ju, Mary Combs, Samuel D Lendle, \\ Jessica M Franklin, Richard Wyss, \\ Sebastian Schneeweiss, Mark J. van der Laan}
\date{}

\maketitle

\begin{abstract}

  The optimal learner for prediction modeling varies depending on the underlying data-generating distribution. Super Learner (SL) is a generic ensemble learning algorithm that uses cross-validation to select among a ``library'' of candidate prediction models. The SL is not restricted to a single prediction model, but uses the strengths of a variety of learning algorithms to adapt to different databases. While the SL has been shown to perform well in a number of settings, it has not been thoroughly evaluated in large electronic healthcare databases that are common in  pharmacoepidemiology and comparative effectiveness research. In this study, we applied and evaluated the performance of the SL in its ability to predict treatment assignment using three electronic healthcare databases. We considered a library of algorithms that consisted of both nonparametric and parametric models. We also considered a novel strategy for prediction modeling that combines the SL with the high-dimensional propensity score (hdPS) variable selection algorithm. Predictive performance  was assessed  using three metrics: the negative log-likelihood, area under the curve (AUC), and time complexity. Results showed that the best individual algorithm, in terms of predictive performance, varied across datasets. The SL was able to adapt to the given dataset and optimize predictive performance relative to any individual learner. Combining the SL with the hdPS was the most consistent prediction method and may be promising for PS estimation and prediction modeling in electronic healthcare databases.

\end{abstract}

\section{Introduction}
\label{s:1}

Traditional approaches to prediction modeling have primarily included parametric models like logistic regression \citep{brookhart2006variable}. While useful in many settings, parametric models require strong assumptions that are not always satisfied in practice. Machine learning methods, including classification trees, boosting, and random forest , have been developed to overcome the limitations of parametric models by requiring assumptions that are less restrictive \citep{hastie2009elements}. Several of these methods have been evaluated for modeling propensity scores and have been shown to perform well in many situations when parametric assumptions are not satisfied  \citep{setoguchi2008evaluating,lee2010improving,westreich2010propensity,wyss2014role}. 
No single prediction algorithm, however, is optimal in every situation and the best performing prediction model will vary across different settings and data structures.

Super Learner (SL) is a general loss-based learning method that has been proposed and analyzed theoretically in  \citep{van2007super}. It is an ensemble learning algorithm that creates a weighted combination of many candidate learners to build the optimal estimator in terms of minimizing a specified loss function. It has been demonstrated that the SL performs asymptotically at least as well as the best choice among the library of candidate algorithms if the library does not contain a correctly specified parametric model; otherwise, it achieves the same rate of convergence as the correctly specified parametric model \citep{van2003unified,dudoit2005asymptotics,vaart2006oracle}.  While the SL has been shown to perform well in a number of settings \citep{van2007super,gruber2015ensemble,rose2016machine}, it's performance has not been thoroughly investigated within large electronic healthcare datasets that are common in pharmacoepidemiology and medical research.  Electronic healthcare datasets based on insurance claims data are different from traditional medical datasets. It is impossible to directly use all of the claims codes as input covariates for supervised learning algorithms, as the number of codes could be larger than the sample size.

In the this study, we compared several statistical and machine learning prediction algorithms  for estimating propensity scores (PS) within three electronic healthcare datasets. We considered a library of algorithms that consisted of both nonparametric and parametric models. We also considered a novel strategy for prediction modeling that  combines the SL with an automated variable selection algorithm for electronic healthcare databases known as the high-dimensional propensity score (hdPS) (discussed later). The predictive performance for each of the methods was assessed using the negative log-likelihood, AUC (i.e., c-statistic or area under the curve), and time complexity. While the goal of the PS is to control for confounding by balancing covariates across treatment groups, in this study we were interested in evaluating the predictive performance of the various PS estimation methods. This study extends previous work that has implemented the SL within electronic healthcare data by proposing and evaluating the novel strategy of combining the SL with the hdPS variable selection algorithm for PS estimation. This study also provides the most extensive evaluation of the SL within healthcare claims data by utilizing three separate healthcare datasets and considering a large set of supervised learning algorithms, including the direct implementation of hdPS generated variables within the supervised algorithms.






\section{Data Sources and Study Cohorts}
\label{S:2}

We used three published healthcare datasets \citep{schneeweiss2009high,ju2016scalable} to assess the performance of the models: the Novel Oral Anticoagulant Prescribing (NOAC) data set, the Nonsteroidal anti-inflammatory drugs (NSAID)
data set and the Vytorin data set. Each dataset consisted of two types of covariates: baseline
covariates  which were selected a priori using expert knowledge, and claims codes. Baseline covariates included demographic variables (e.g. age, sex, census region and race) and other predefined
covariates that were selected a priori using expert knowledge. Claims codes included information on diagnostic, drug, and procedural insurance claims for individuals within the healthcare databases.

\subsection{Novel Oral Anticoagulant (NOAC) Study}

The NOAC data set was generated to track a cohort of new-users of oral anticoagulants to study the comparative safety and effectiveness of warfarin versus dabigatran in preventing stroke. 
Data were collected by United Healthcare  between October, 2009 and December, 2012. The dataset includes 18,447 observations, 60 pre-defined baseline covariates and 23,531 unique claims codes.  Each claims code within the dataset records the number of times that specific code occurred for each patient within a pre-specified baseline period prior to initiating treatment. The claims code covariates fall into four categories, or "data dimensions": inpatient diagnoses, outpatient diagnoses, inpatient procedures and outpatient procedures. For example, if a patient has a value of 2 for the variable "pxop\_V5260", then the patient received the outpatient procedure coded as V5260 twice during the pre-specified baseline period prior to treatment initiation.

\subsection{Nonsteroidal anti-inflammatory drugs (NSAID) Study}

The NSAID dataset was constructed to compare new-users of a selective COX-2 inhibitor versus a nonselective NSAID in the risk of GI bleed. The observations were drawn from a population of patients aged 65 years and older who were enrolled in both Medicare and the Pennsylvania Pharmaceutical Assistance Contract for the Elderly (PACE) programs between 1995 and 2002.  The dataset consists of 49,653 observations, with 22 pre-defined baseline covariates and 9,470 unique claims codes \citep{schneeweiss2009high}. The claims codes fall into eight data dimensions: prescription drugs, ambulatory diagnoses, hospital diagnoses, nursing home diagnoses, ambulatory procedures, hospital procedures, doctor diagnoses and doctor procedures.

\subsection{Vytorin Study}

The Vytorin dataset was generated to track a cohort of new-users of Vytorin and high-intensity statin therapies. The data were collected to study the effects of these medications on  the combined outcome, myocardial infarction, stroke and death. The dataset includes all United Healthcare patients between January 1, 2003 and December 31, 2012, who were 65 years of age or older on the day of entry into the study cohort \citep{schneeweiss2012supplementing}. 
The dataset consists of 148,327 individuals, 67 pre-defined  baseline covariates and 15,010 unique claims codes. The claims code covariates fall into five data dimensions: ambulatory diagnoses, ambulatory procedures, prescription drugs, hospital diagnoses and hospital procedures.

\section{Methods}
\label{S:3}

In this paper, we used R (version 3.2.2) for the data analysis. For each dataset, we randomly selected $80\%$ of the data as the training set and the rest as the testing set. We centered and scaled each of the covariates as some algorithms are sensitive to the magnitude of the covariates. We conducted model fitting and selection only on the training set, and assessed the goodness of fit for all of the models on  the testing set to ensure objective measures of prediction reliability.

\subsection{The high-dimensional propensity score algorithm}

The high-dimensional propensity score (hdPS) is an automated variable selection algorithm that is designed to identify confounding variables within electronic healthcare databases.  Healthcare claims databases contain multiple data dimensions, where each dimension represents a different aspect of healthcare utilization (e.g., outpatient procedures, inpatient procedures, medication claims, etc.). When implementing the hdPS, the investigator first specifies how many variables to consider within each data dimension. Following the notation of \citep{schneeweiss2009high} we let $n$ represent this number. For example, if $n=200$ and there are 3 data dimensions, then the hdPS will consider 600 codes.

For each of these 600 codes, the hdPS then creates three binary variables labeled “frequent”, “sporadic”, and “once” based on the frequency of occurrence for each code during a covariate assessment period prior to the initiation of exposure. In this example, there are now a total of 1,800 binary variables. The hdPS then ranks each variable based on its potential for bias using the Bross formula \citep{bross1966spurious,schneeweiss2009high}. Based on this ordering, investigators then specify the number of variables to include in the hdPS model, which is represented by $k$. A detailed description of the hdPS is provided by \citet{schneeweiss2009high}.

\subsection{Machine Learning Algorithm Library}

We evaluated the predictive performance of a variety of machine learning algorithms that are available within the caret package (version 6.0) in the R programming environment  \citep{kuhn2008caret,kuhn2014caret} . Due to computational constraints, we screened the available algorithms to only include those that were computationally less intensive. A list of the chosen algorithms is provided in the Web Appendix. 

Because of the large size of the data, we used leave group out (LGO) cross-validation instead of $V$-fold cross-validation to select the tuning parameters  for each individual algorithm. We randomly selected $90\%$ of the training  data for model training and $10\%$ of the training data for model tuning and selection. For clarity, we refer to these subsets of the training data as the LGO training set and the LGO validation set, respectively. After the tuning parameters were selected, we fitted the selected models on the whole training set, and assessed the predictive performance for each of the models on the testing set. See the appendix for more details of the base learners.

\begin{figure}[H]
  \begin{center}
    \includegraphics[width=0.7\textwidth]{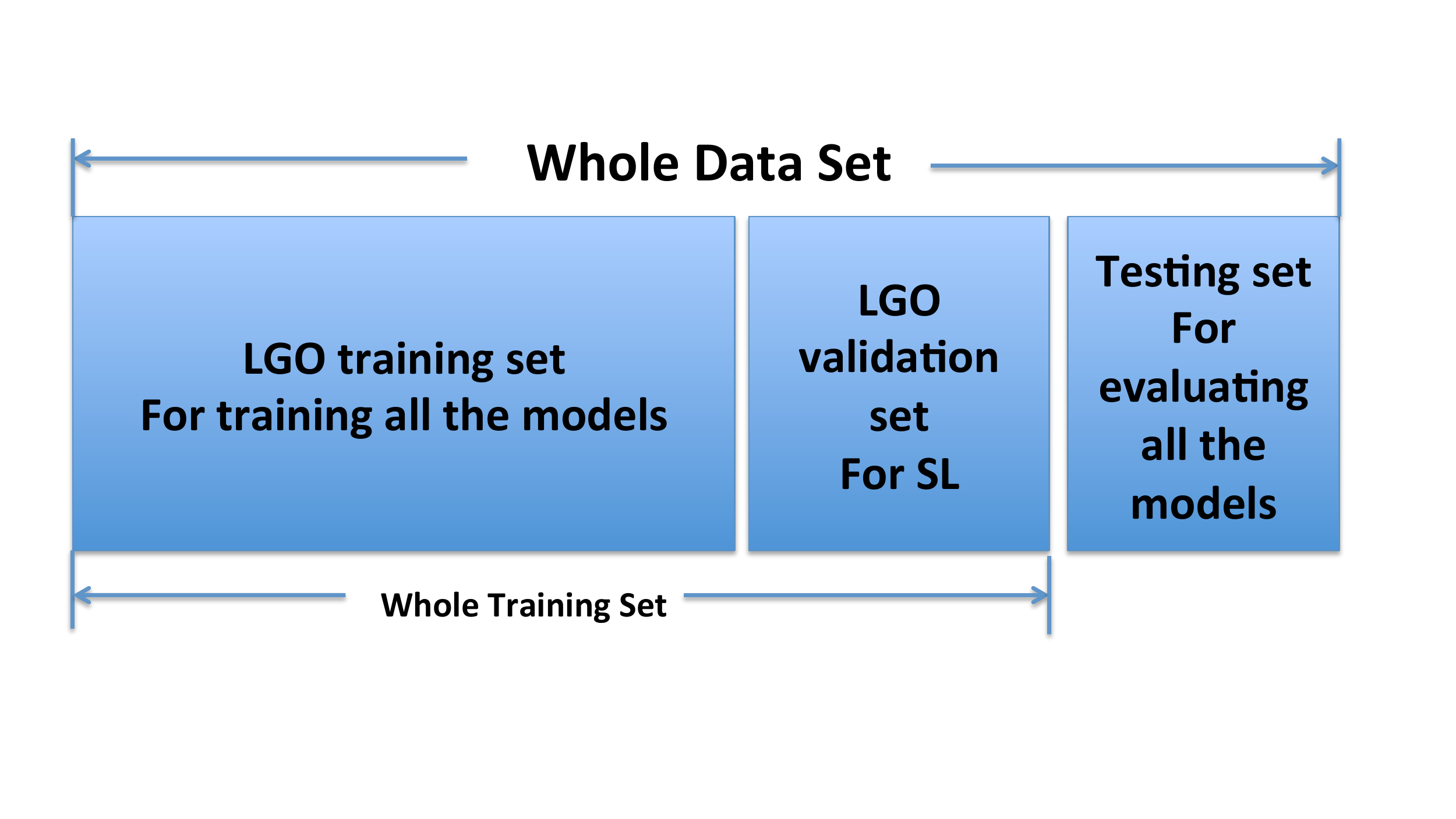}
    \caption{The split of dataset}
    \label{fig:split}
  \end{center}
\end{figure}


\subsection{Super Learner}

Super Learner (SL) is a method for selecting an optimal prediction algorithm from a set of user-specified prediction models. The SL relies on the choice of a loss function (negative log-likelihood in the present study) and the choice of a library of candidate algorithms.  The SL then compares the performance of the candidate algorithms using  V-fold cross-validation: 
for each candidate algorithm, SL averages the estimated risks across the validation sets, resulting in the so-called cross-validated risk. Cross-validated risk estimates are then used to compute the best  weighted linear convex combination of the candidate learners with the smallest estimated risk.  This weighted combination  is then applied to the full study data to produce a new set of predicted values and is referred to as the SL estimator \citep{van2007super,polley2010super}. \citet{benkeser2016online} further proposed an online-version of SL for streaming big data.

Due to  computational constraints, in this study,  we used LGO validation instead of V-fold cross-validation  when implementing the SL algorithm. We first fitted every candidate algorithm on the LGO training set, then computed the best weighted combination for the SL on the LGO validation set. This variation of the SL algorithm is known as the sample split SL algorithm. We used the SL package in R (Version: 2.0-15) to evaluate the predictive performance of three SL estimators:

\begin{description}
  
\item[SL1] Included only pre-defined  baseline variables with all 23 of the previously identified traditional machine learning algorithms in the SL library.

\item[SL2] Identical to SL1, but included the hdPS algorithms with various tuning parameters. Note that in SL2, only the hdPS algorithms had access to the claims code variables.

\item[SL3] Identical to SL1, but included both pre-defined baseline variables and hdPS generated variables within the traditional learning algorithms. Based on the performance of each individual hdPS algorithms, a fixed pair of hdPS tuning parameters was selected in order to find the optimal ensemble of all candidate algorithms that were  fitted on the same set of variables.

\end{description}

\begin{table}[H]
  \centering
  \scalebox{0.85}{
  \begin{tabular}{|cp{6cm}p{8cm}|}
    \hline
    \multicolumn{1}{|>{\centering\arraybackslash}m{1.3cm}}{Super Learner} 
    & \multicolumn{1}{>{\centering\arraybackslash}m{5.5cm}}{Libray} 
    & \multicolumn{1}{>{\centering\arraybackslash}m{7cm}|}{Covariates}\\ \hline

    SL1 & All machine learning algorithms  & Only baseline covariates.  \\ \hline
    SL2 & All machine learning algorithms and the hdPS 
    algorithm & Baseline covariates; Only the hdPS algorithm utilizes the claims codes. \\ \hline 
    SL3 & All machine learning algorithms & Baseline
    covariates and hdPS covariates generated from claims codes. \\ \hline
  \end{tabular}
  }
  \caption{Details of the three Super Learners considered.}
  \label{table:SLs}
\end{table}

\subsection{Performance Metrics}

We used three criteria to evaluate the prediction algorithms: computing time, negative log-likelihood, and area under the curve (AUC). In statistics, a receiver operating characteristic (ROC), or ROC curve, is a plot that illustrates the performance of a binary classifier system as its discrimination threshold is varied. The curve is created by plotting the true positive rate against the false positive rate at various threshold settings. The AUC is then computed as the area under the ROC curve.

For both computation time and negative log-likelihood, smaller values indicate better performance, whereas for AUC the better classifier achieves greater values \citep{hanley1982meaning}. Compared to the error rate, the AUC is a better assessment of performance for the unbalanced classification problem.

\section{Results}
\label{S:4}
\subsection{Using the hdPS prediction algorithm with Super Learner}
\subsubsection{Computation Times}

\begin{figure}[H]
  \centering
  \begin{subfigure}[b]{0.7\textwidth}
    \includegraphics[width=\textwidth]{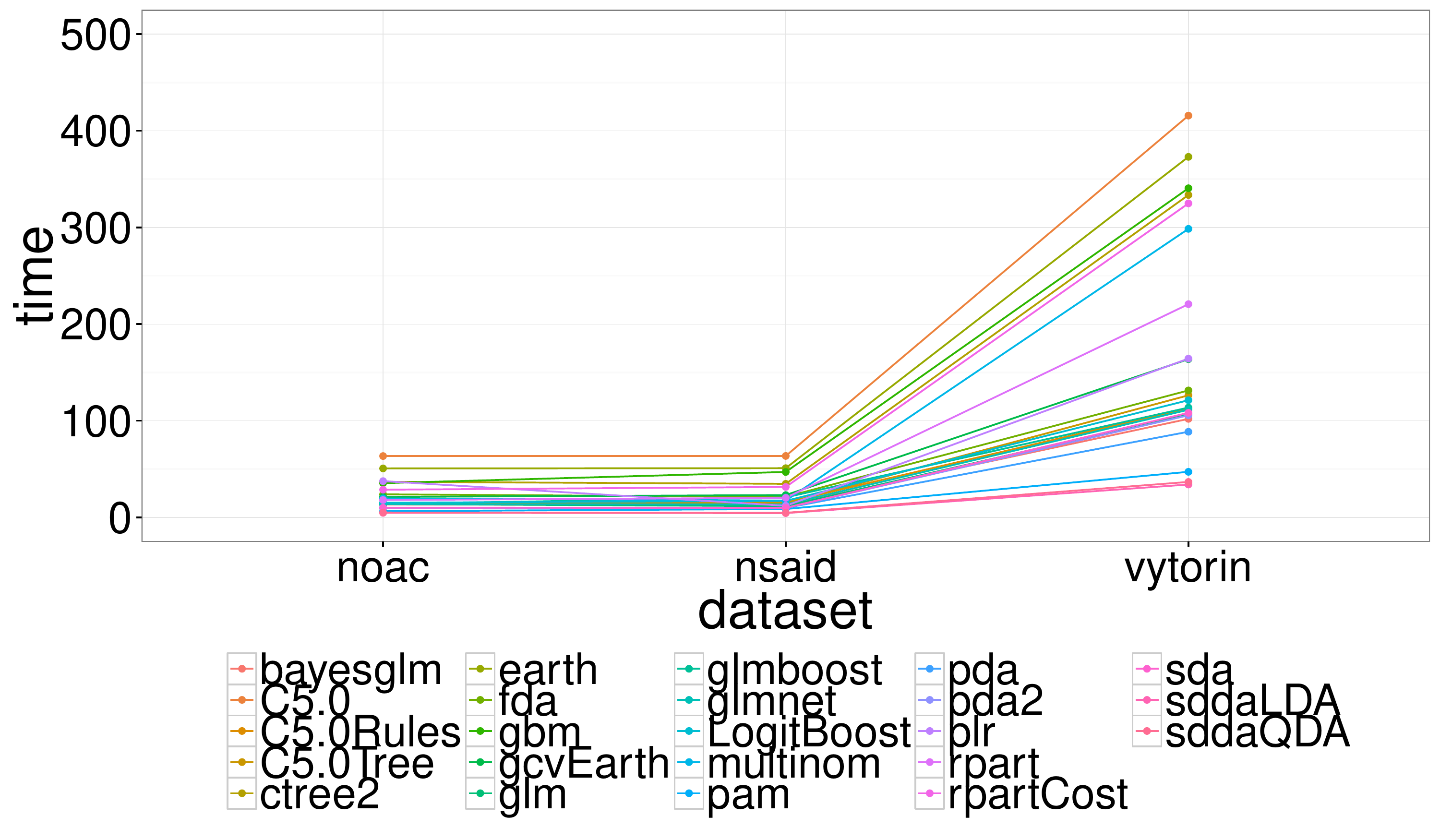}
    \caption{Running time (in second)for the 23 individual machine learning algorithms with no Super Learner.}
    \label{time_all}
  \end{subfigure}
  \begin{subfigure}[b]{0.7\textwidth}
    \includegraphics[width=\textwidth]{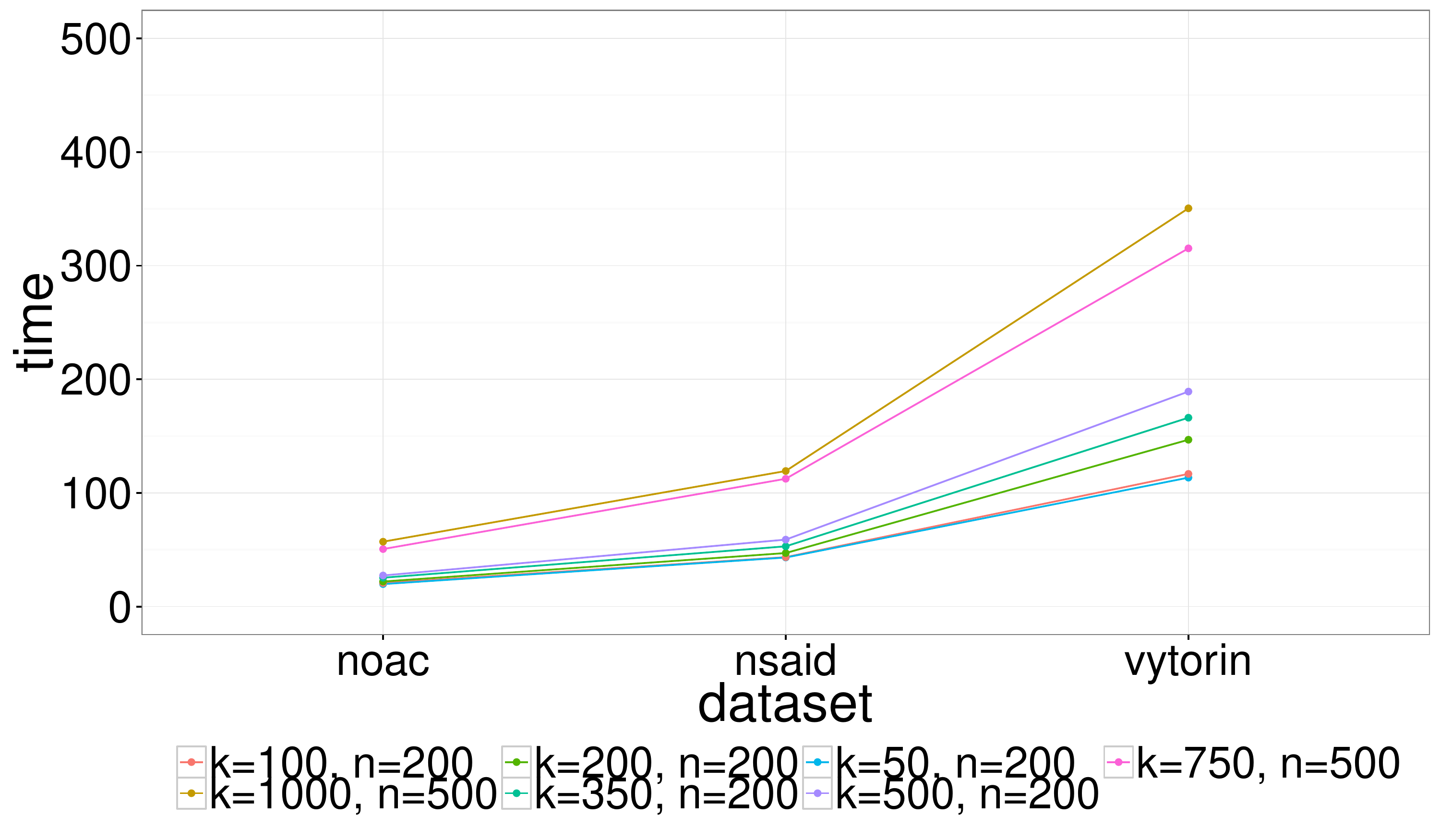}
    \caption{Running time for the hdPS algorithms varying the parameter $k$ from $50$ to $750$ for $n=200$, and $n=500$.}
    \label{time_hdps}
  \end{subfigure}
  \caption{Running times for individual machine learning and hdPS algorithms without Super Learner. The y-axis is in log scale.}
  \label{time}
\end{figure}

Figure \ref{time} shows the running time for the 23 individual machine learning
algorithms and the hdPS algorithm across all three datasets without the use of Super
Learner. Running time is measured in seconds. Figure \ref{time_all} shows the running 
time for the machine learning algorithms that only use baseline covariates.
Figure \ref{time_hdps} shows the running time for the hdPS algorithm at varying values of the tuning  parameters $k$ and $n$. Recall $n$ represents the number of variables that the hdPS algorithm considers within each data dimension and $k$ represents the total number of variables that are selected or included in the final hdPS model as discussed previously. The running time is sensitive to $n$,  while less sensitive to $k$. This suggests that most of the running  time for the hdPS is spent generating and screening covariates. The running time for the hdPS algorithm is generally around the median of all the running times of the machine learning algorithms that included only baseline covariates. Here we only compared the running time for each pair of parameters for hdPS. It is worth noting that the variable creation and ranking only has to be done once for each value of $n$. Modifying values of $k$ just means taking different numbers of variables from a list and refitting the logistic regression. 

The running time of SL is not placed in the figures. SL with baseline covariates takes just over twice as long as the sum of the running time for each individual algorithm in its library: SL  splits data into training and validation sets, fits the base learners on the training set, finds weights based the on the validation set, and finally retrains the model on the whole set. In other words, Super Learner fits every single algorithm twice, with additional processing time for computing the weights. Therefore, the running time will be about twice the sum of its constituent algorithms, which is what we see in this study (see Table \ref{table:SL_time}).

\begin{table}[H]
  \centering
  \begin{tabular}{|crc|}
    \hline
    \multicolumn{1}{|>{\centering\arraybackslash}m{23mm}}{Data Set} 
 	  & \multicolumn{1}{>{\centering\arraybackslash}m{65mm}}{Algorithm} 
    & \multicolumn{1}{>{\centering\arraybackslash}m{35mm}|}{Processing Time (seconds)}\\ \hline

    NOAC &Sum of machine learning algorithms  & 481.13  \\ 
    &Sum of hdPS  algorithms& 222.87  \\ 
    & Super Learner 1  & 1035.43  \\ 
    & Super Learner 2  &  1636.48 \\ \hline

    NSAID & Sum of machine learning algorithms&   476.09   \\ 
    & Sum of hdPS algorithms&   477.32   \\ 
    & Super Learner 1 &    1101.84  \\ 
    & Super Learner 2 &    2075.05  \\ \hline
    VYTORIN & Sum of machine learning algorithms & 3982.03 \\ 
    & Sum of hdPS algorithms & 1398.01 \\ 
    & Super Learner 1 & 9165.93 \\ 
    & Super Learner 2 & 15743.89 \\ \hline
  \end{tabular}
  \caption{Running time of the machine learning algorithms, the hdPS algorithms, and 
	  Super Learners 1 and 2. Twice the sum of the
	  running time of the machine learning algorithms is comparable to the running time of
	  Super Learner 1 and twice the sum of the running times of both the machine learning algorithms and 
	  the hdPS algorithms is comparable to the running time of Super Learner 2.}
  \label{table:SL_time}
\end{table}


\subsubsection{Negative log-likelihood}

\begin{figure}[H]
  \centering
  \begin{subfigure}[t]{0.7\textwidth}
    \includegraphics[width=\textwidth]{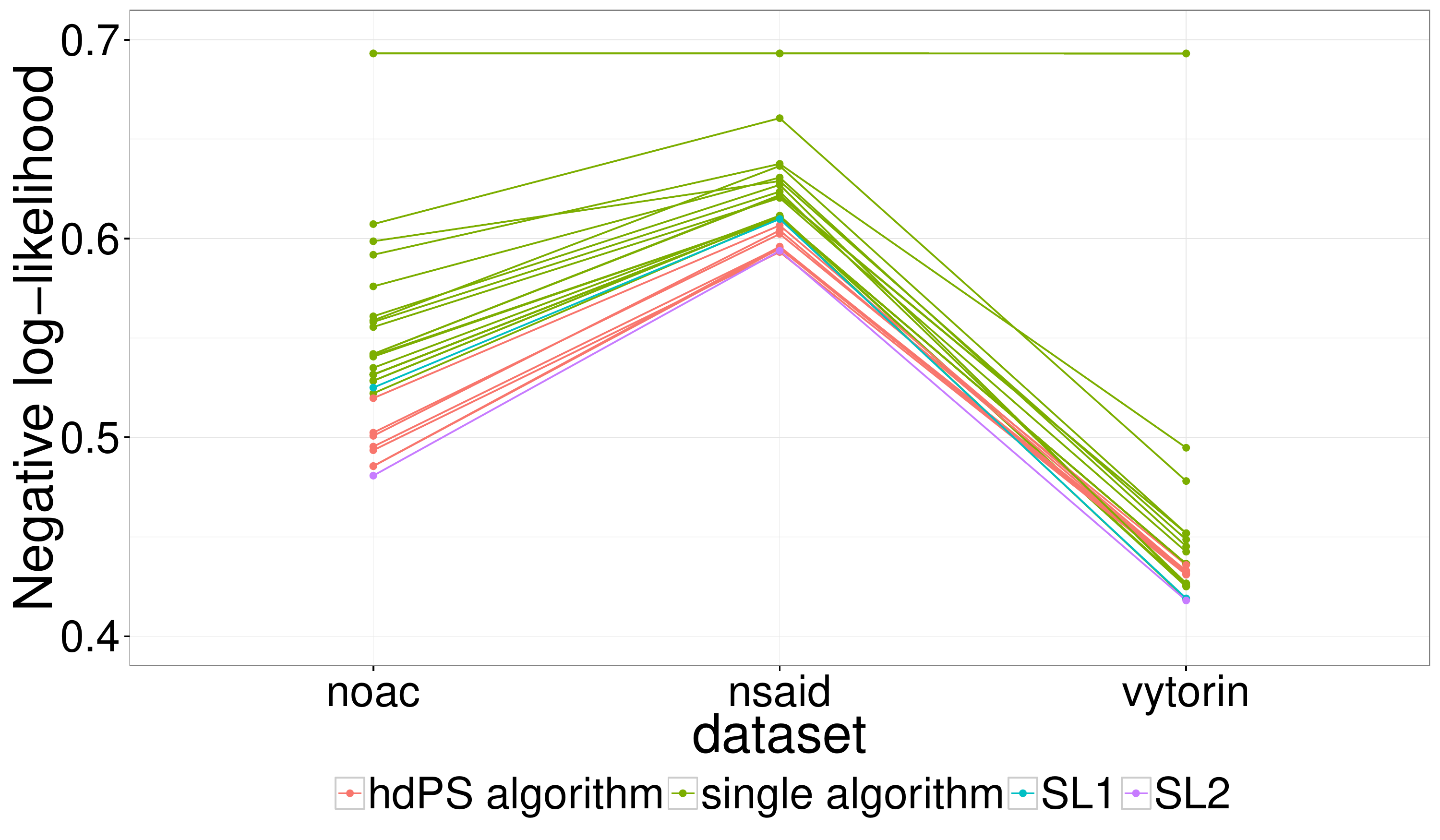}
    \caption{Negative log-likelihood for SL1, SL2, the hdPS algorithm, and the 23 machine learnng algorithms.}
    \label{likelihood_all}
  \end{subfigure}
  \begin{subfigure}[t]{0.7\textwidth}
    \includegraphics[width=\textwidth]{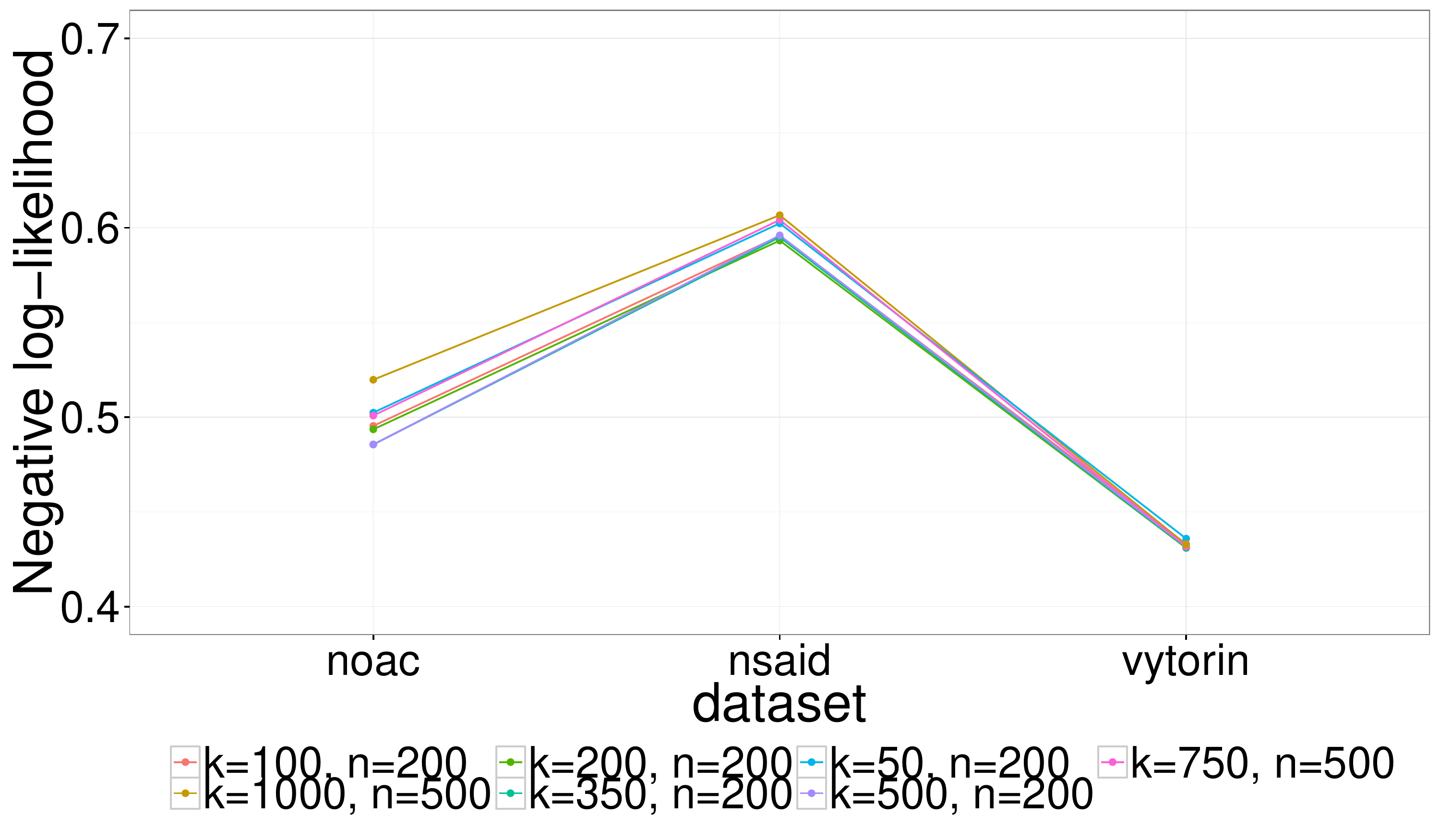}
    \caption{Negative log-likelihood for the hdPS algorithm, varying the parameter $k$ from $50$ to $750$ for $n=200$, and $n=500$.}
    \label{likelihood_hdps}
  \end{subfigure}
  \caption{The negative log-likelihood for SL1, SL2, the hdPS algorithm, and the 23 machine learning algorithms.}\label{likelihood}
\end{figure}

Figure \ref{likelihood_all}  shows the negative
log-likelihood for Super Learners 1 and 2, and each of the 23 machine learning algorithms (with
only baseline covariates) .  Figure \ref{likelihood_hdps} shows the negative
log-likelihood for  hdPS algorithms with varying tuning parameters, $n$ and $k$.

For these examples, figure \ref{likelihood_hdps} shows that the performance of the hdPS, in terms of reducing the negative log-likelihood, is not sensitive to either $n$ or $k$. Figure \ref{likelihood} further shows that the hdPS generally outperforms the majority of the individual machine learning algorithms within the library, as it takes advantage of the extra information from the claims codes. However, in the Vytorin data set, there are still some machine learning algorithms which perform slightly better than the hdPS with respect to the negative log-likelihood.

Figure \ref{likelihood_all} shows that the SL (without hdPS) outperforms all the other
individual algorithms  in terms of reducing the negative log-likelihood. 
The figures further show that the predictive performance of the SL improves when the hdPS algorithm is included within the SL library of candidate algorithms. With the help of the hdPS, the SL results in the greatest reduction in the negative log-likelihood when compared to all of the individual prediction algorithms (including the hdPS itself).

\subsubsection{AUC}

\begin{figure}[H]
  \centering
  \begin{subfigure}[t]{0.7\textwidth}
    \includegraphics[width=\textwidth]{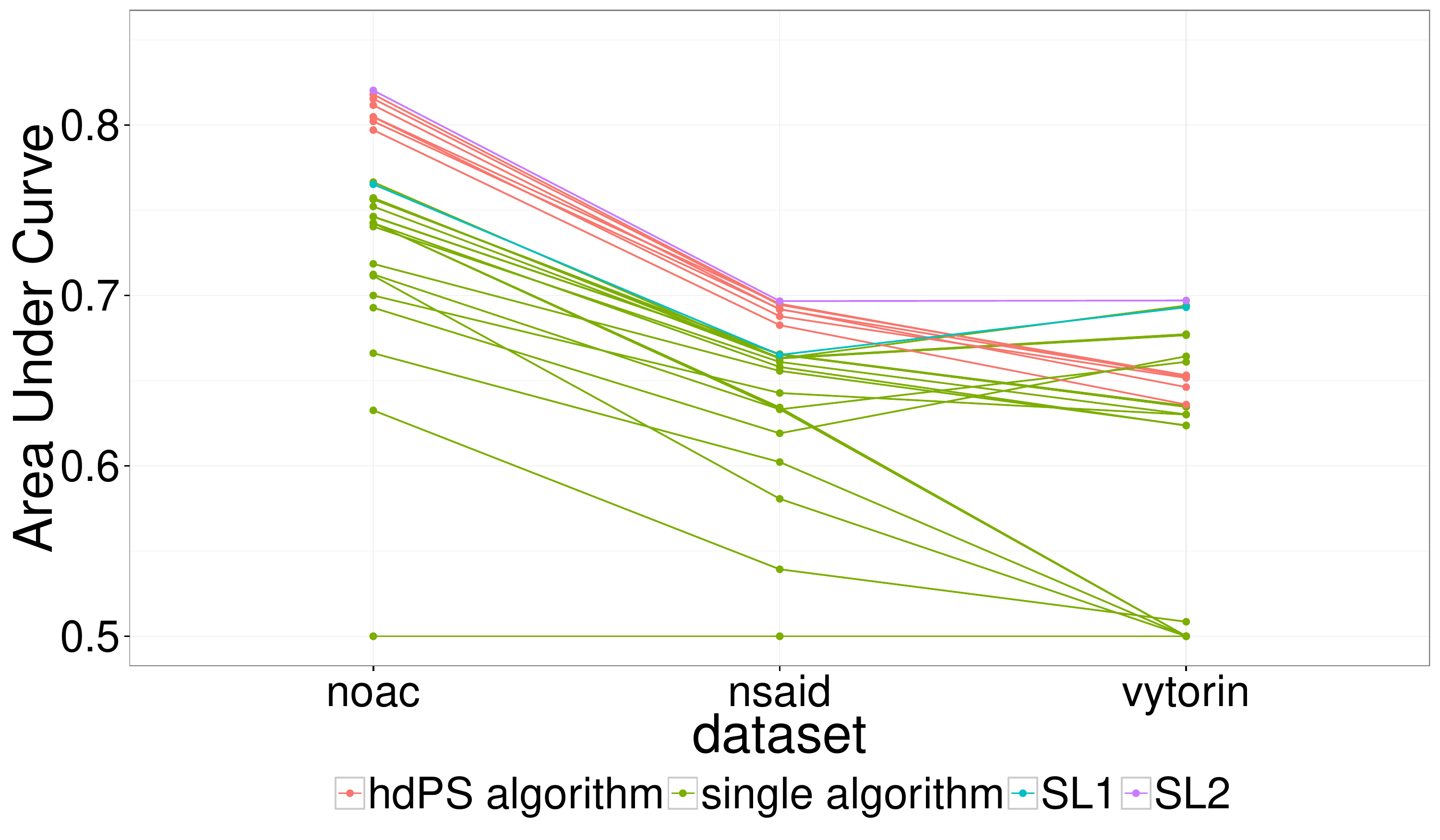}
    \caption{AUC of SL1, SL2, the hdPS algorithm, and the 23 machine learnng algorithms.}
    \label{AUC_all}
  \end{subfigure}
  \begin{subfigure}[t]{0.7\textwidth}
    \includegraphics[width=\textwidth]{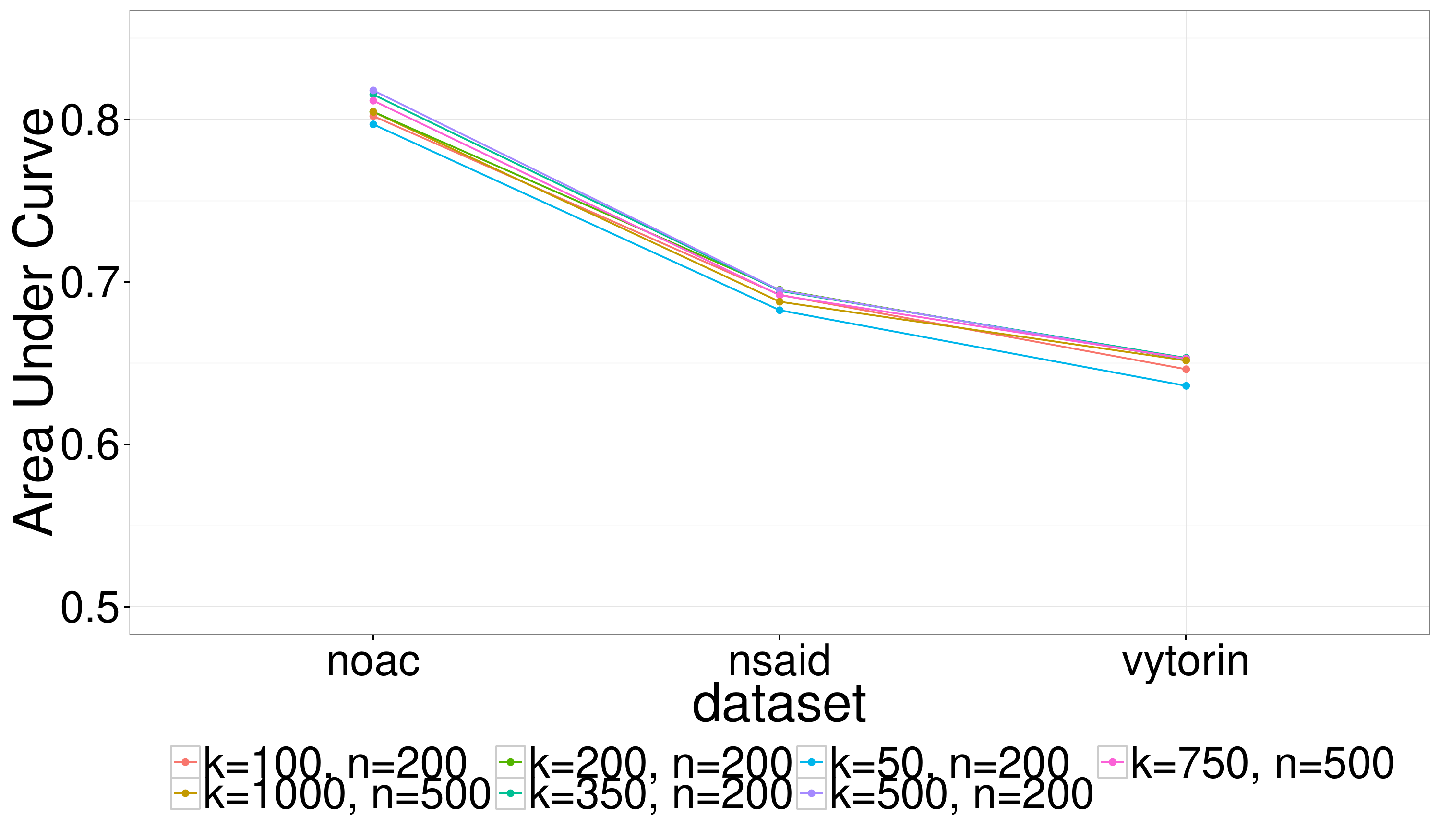}
    \caption{AUC for the hdPS algorithm, varying the parameter $k$ from $50$ to $750$ for $n=200$, and $n=500$.}
    \label{AUC_hdps}
  \end{subfigure}
  \caption{The area under the ROC curve (AUC) for for Super Learners 1 and 2, the hdPS algorithm, and each of the 23 machine learning algorithms.}
  \label{AUC}
\end{figure}

The SL uses loss-based cross-validation to select the optimal combination of individual algorithms. Since the negative log-likelihood was selected as the loss function when running the SL algorithm, it is not surprising that it outperforms other algorithms with respect to the negative log-likelihood. As PS estimation can be considered a binary
classification problem, we can also use the Area Under the Curve (AUC) to compare 
performance across algorithms.
Binary classification is typically determined by setting a threshold. 
As the threshold varies for a given classifier we can achieve different true
positive rates (TPR) and false positive rates (FPR). A Receiver Operator Curve (ROC) space is
defined by FPR and TPR as the x- and y-axes respectively, to depict
the trade-off between true positives (benefits) and false
positives (costs) at various classification thresholds. We then draw the ROC curve of 
TPR and FPR for each model and calculate the AUC. The upper bound for
a perfect classifier is 1 while a naive random guess would achieve about 0.5.

In Figure \ref{AUC_all}, we compare the performance of Super Learners 1 and 2, the hdPS algorithm, and each of the 23 machine learning algorithms. Although we optimized  Super Learners with respect to the negative log-likelihood loss function, SL1 and SL2 showed good performance with respect to the AUC; In the NOAC and  NSAID data sets, the hdPS algorithms outperformed SL1, in terms of maximizing the AUC, but SL1 (with only baseline variables) achieved a higher value for AUC,  compared to each of the individual  machine learning algorithms in its library. In the VYTORIN data set, SL1 outperformed hdPS algorithms with respect to AUC, even though the hdPS algorithms use the additional claims data. Table \ref{table:AUC} shows that, in all three data sets, the SL3 achieved higher AUC  values compared to all the other algorithms, including hdPS and SL1.

\begin{table}
  \centering
  \begin{tabular}{|l|l|l|l|}
    \hline
    data    & SL1     & SL2     & best hdPS (parameter k/n)  \\ \hline
    noac    & 0.7652  & 0.8203  & 0.8179 (500/200)           \\ \hline
    nsaid   & 0.6651  & 0.6967  & 0.6948 (500/200)           \\ \hline
    vytorin & 0.6931  & 0.6970  & 0.6527 (750/500)           \\ \hline
  \end{tabular}
  \caption{Comparison of AUC for SL1, SL2 and best hdPS across three data sets. The best hdPS for noac is k = 500, n = 200, and for nsaid is k = 500, n = 200, for vytorin is k = 750, n = 500.}
  \label{table:AUC}
\end{table}

\subsection{Using the hdPS screening method with Super Learner}

In the previous sections, we compared machine learning algorithms that were limited to only baseline covariates with the hdPS algorithms across two different measures of performance (negative log-likelihood and AUC). The results showed that including the hdPS algorithm within the SL library improved the predictive  performance. In this section, we combined the information that is contained within the claims codes via the hdPS screening method with the machine learning algorithms.

We first used the hdPS screening method (with tuning parameters  $n = 200,k = 500$) to generate and screen the hdPS covariates. We then combined these hdPS covariates with the pre-defined baseline covariates to generate augmented datasets for each of the three datasets under consideration.
We built a SL library that included each of the 23 individual machine learning algorithms, fitted on both baseline and hdPS generated covariates. Note that, as the original hdPS method uses logistic regression for prediction, it can be considered a special case of LASSO (with $\lambda=0$). For simplicity, we use ``Single algorithm'' to denote the conventional machine learning algorithm with only baseline covariates, and  ``Single algorithm*''  to denote  the single machine learning algorithm in the library.

\begin{figure}[H]
  \centering
  \begin{subfigure}[t]{0.7\textwidth}
    \includegraphics[width=\textwidth]{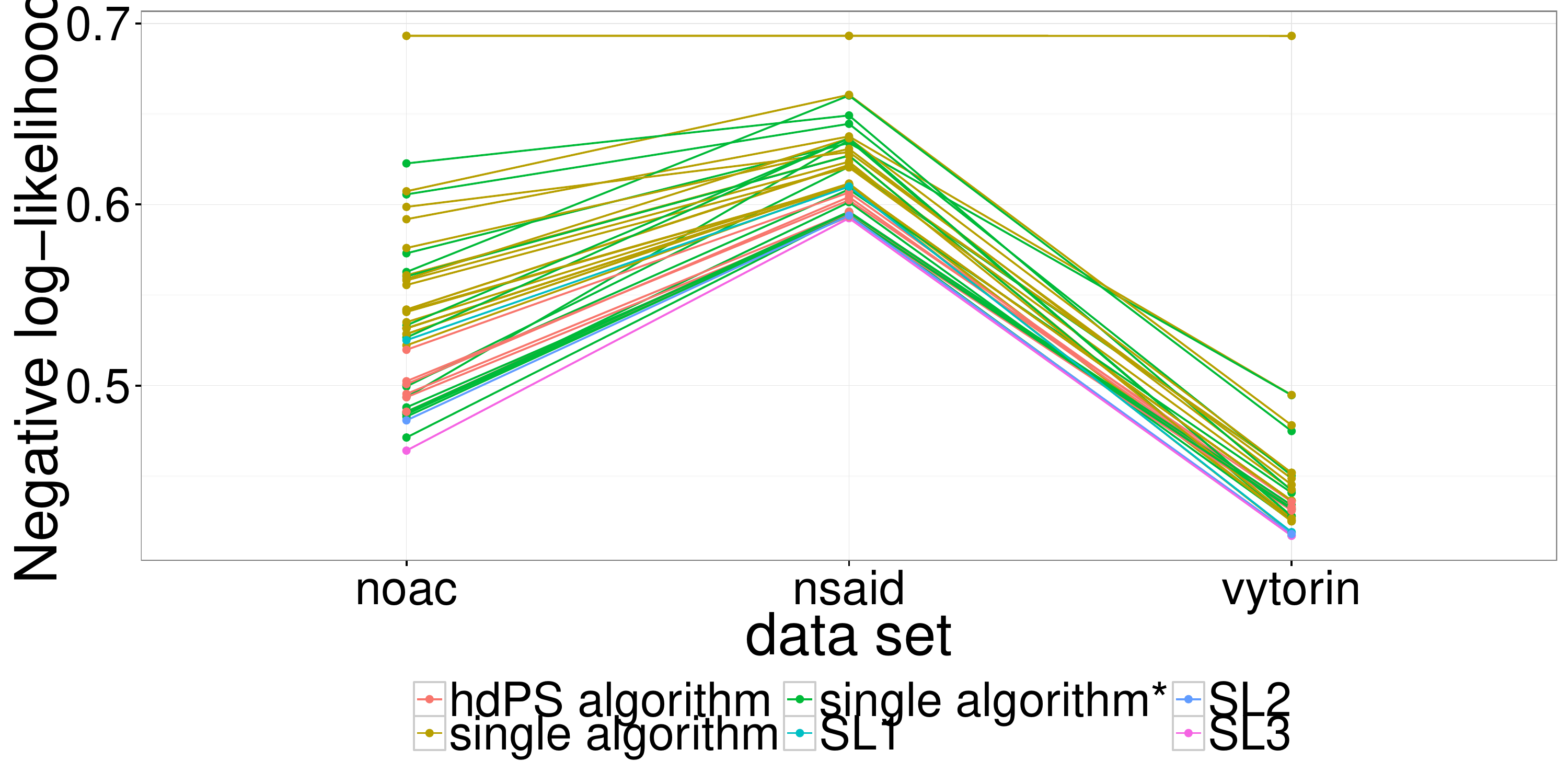}
    \caption{Negative log-likelihood}
    \label{likelihood_screen}
  \end{subfigure}
  \begin{subfigure}[t]{0.7\textwidth}
    \includegraphics[width=\textwidth]{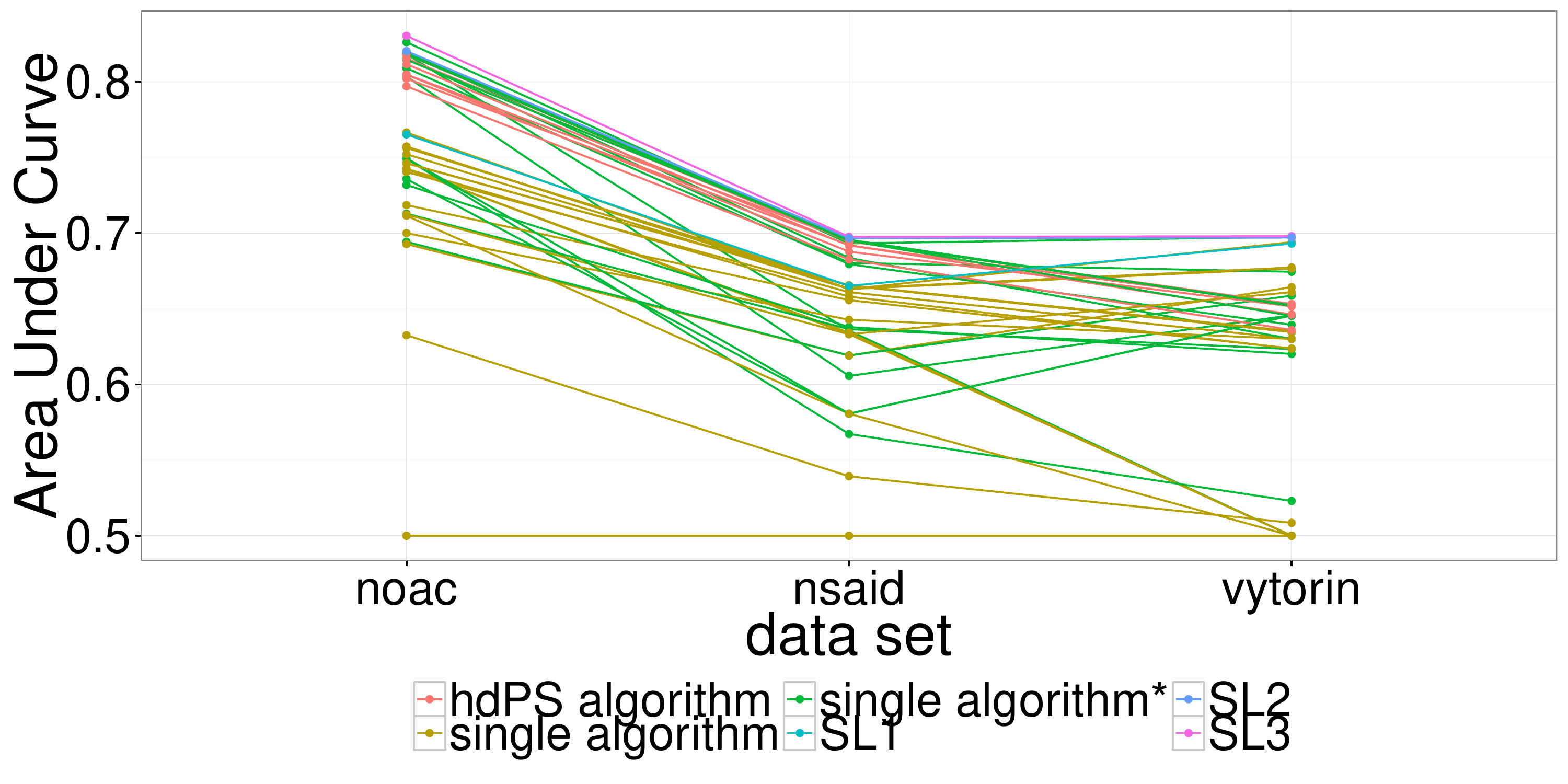}
    \caption{AUC }
    \label{AUC_screen}
  \end{subfigure}
  
  \caption{Negative log-likelihood and AUC of SL1, SL2, and SL3, compared with each of the single machine learning algorithms with and without using hdPS covariates. We use ``Single algorithm'' to denote the conventional machine learning algorithm with only baseline covariates, and  ``Single algorithm*''  to denote  the single machine learning algorithm in the library.}
  \label{fig:screened}
\end{figure}

For convenience, we differentiate Super Learners 1, 2 and 3 by their 
algorithm libraries: machine learning algorithms with only baseline covariates, 
augmenting this library with hdPS, and only the machine learning algorithms but with
both baseline and hdPS screened covariates (see Table \ref{table:SLs}).

Figures \ref{fig:screened} compares the negative log-likelihood and AUC, respectively, of all three Super Learners and machine learning algorithms. Figure \ref{fig:screened} shows that the
performance of all algorithms increases after including the hdPS generated variables. Figure \ref{fig:screened} further shows that  SL3 performs slightly better than SL2, but the difference is small.

\begin{center}
  \begin{table}[H]
    \scalebox{0.85}{
    \begin{tabular}{|ccccc|}
      \hline
      Data set & Performance Metric & Super Learner 1 & Super Learner2 & Super Learner 3  \\ \hline
      NOAC &  & 0.7652 & 0.8203 & 0.8304 \\ 
      NSAID & AUC & 0.6651 & 0.6967 & 0.6975 \\ 
      VYTORIN & & 0.6931 & 0.6970 & 0.698 \\ \hline

      NOAC & & 0.5251 & 0.4808 & 0.4641 \\ 
      NSAID & Negative Log-likelihood  & 0.6099 & 0.5939 & 0.5924 \\ 
      VYTORIN & & 0.4191 & 0.4180 & 0.4171 \\ \hline
    \end{tabular}
    }
    \caption{Performance as measured by AUC and negative log-likelihood for the three Super Learners with the following libraries: machine learning algorithms with only baseline covariates, augmenting this library with hdPS, and only the machine learning algorithms but with
      both baseline and hdPS screened covariates. (See Table \ref{table:SLs}).}
    \label{table:comparison}
  \end{table}
\end{center}

Table \ref{table:comparison} shows that performances were improved from SL 1 to 2 and from 2 to 3. The differences in the AUC and in the negative log-likelihood between SL1 and 2 are large, while these differences between SL2 and 3 are small. This suggests two things: First, the prediction step in the hdPS algorithm (logistic regression) works well in these datasets: it performs approximately as well as the best individual machine learning algorithm in the library for SL 3. Second, the hdPS screened covariates make the PS estimation more flexible; using SL we can easily develop different models/algorithms which incorporate the covariate screening method from the hdPS.

\subsection{Weights of Individual Algorithms in Super Learners 1 and 2}

  \begin{table}[H]
    \centering
    \scalebox{0.7}{
    \begin{tabular}{|clc|}
      \hline
      \textbf{Data Set} & \textbf{Algorithms Selected for SL1} & \textbf{Weight} \\
      \hline
      NOAC & SL.caret.bayesglm\_All & 0.30 \\
      & SL.caret.C5.0\_All & 0.11 \\
      & SL.caret.C5.0Tree\_All & 0.11 \\
      & SL.caret.gbm\_All & 0.39 \\
      & SL.caret.glm\_All & 0.01 \\
      & SL.caret.pda2\_All & 0.07 \\
      & SL.caret.plr\_All & 0.01 \\\hline
      NSAID & SL.caret.C5.0\_All & 0.06 \\
      & SL.caret.C5.0Rules\_All & 0.01 \\
      & SL.caret.C5.0Tree\_All & 0.06 \\
      & SL.caret.ctree2\_All & 0.01 \\
      & SL.caret.gbm\_All & 0.52 \\
      & SL.caret.glm\_All & 0.35 \\\hline
      VYTORIN & SL.caret.gbm\_All & 0.93 \\
      & SL.caret.multinom\_All & 0.07 \\
      \hline
      \hline
      \textbf{Data Set} & \textbf{Algorithms Selected for SL2} & \textbf{Weight}\\
      \hline
      NOAC & SL.caret.C5.0\_screen.baseline & 0.03 \\
      & SL.caret.C5.0Tree\_screen.baseline & 0.03 \\
      & SL.caret.earth\_screen.baseline & 0.05 \\
      & SL.caret.gcvEarth\_screen.baseline & 0.05 \\
      & SL.caret.pda2\_screen.baseline & 0.02 \\
      & SL.caret.rpart\_screen.baseline & 0.04 \\
      & SL.caret.rpartCost\_screen.baseline & 0.04 \\
      & SL.caret.sddaLDA\_screen.baseline & 0.03 \\
      & SL.caret.sddaQDA\_screen.baseline & 0.03 \\
      & SL.hdps.100\_All & 0.00 \\
      & SL.hdps.350\_All & 0.48 \\
      & SL.hdps.500\_All & 0.19 \\\hline
      NSAID & SL.caret.gbm\_screen.baseline & 0.24 \\
      & SL.caret.sddaLDA\_screen.baseline & 0.03 \\
      & SL.caret.sddaQDA\_screen.baseline & 0.03 \\
      & SL.hdps.100\_All & 0.25 \\
      & SL.hdps.200\_All & 0.21 \\
      & SL.hdps.500\_All & 0.01 \\
      & SL.hdps.1000\_All & 0.23 \\\hline
      VYTORIN & SL.caret.C5.0Rules\_screen.baseline & 0.01 \\
      & SL.caret.gbm\_screen.baseline & 0.71 \\
      & SL.hdps.350\_All & 0.07 \\
      & SL.hdps.750\_All & 0.04 \\
      & SL.hdps.1000\_All & 0.17 \\
      \hline
    \end{tabular}
    }
    \caption{Non-zero weights of individual algorithms in Super Learners 1 and 2 across all three data sets.}
    \label{weight}
  \end{table}

SL produces an optimal ensemble learning algorithm, i.e. a weighted combination of the candidate learners in its library. Table \ref{weight} shows the weights for all the non-zero weighted algorithms included in the data-set-specific ensemble learner generated by SL 1 and 2. Table \ref{weight} shows that for all the three data sets, the gradient boosting algorithm (gbm) has the highest weight.  It is also interesting to note that across the different data sets the hdPS algorithms have very different weights. In the NOAC and NSAID datasets, the hdPS algorithms play a dominating role: hdPS algorithms occupy more than 50\% of the weight. However in the VYTORIN dataset, boosting  plays the most important
role, with a weight of 0.71.

\section{Discussion}

\begin{table}[H]
  \centering
  \scalebox{0.7}{
    \begin{tabular}{|ccccccc|}
      \hline

      \multicolumn{1}{|>{\centering\arraybackslash}m{2cm}}{Data Set} 
      & \multicolumn{1}{>{\centering\arraybackslash}m{3cm}}{Method} 
      & \multicolumn{1}{>{\centering\arraybackslash}m{2cm}}{Negative Log Likelihood}
      & \multicolumn{1}{>{\centering\arraybackslash}m{1cm}}{AUC}
      & \multicolumn{1}{>{\centering\arraybackslash}m{2cm}}{Negative Log Likelihood (Train)}
      & \multicolumn{1}{>{\centering\arraybackslash}m{1.5cm}}{AUC (Train)}
      & \multicolumn{1}{>{\centering\arraybackslash}m{2cm}|}{Processing Time (Seconds)}\\ \hline

      NOAH & k=50, n=200 & 0.50 & 0.80 & 0.51 & 0.79 & 19.77 \\
      & k=100, n=200 & 0.50 & 0.80 & 0.50 & 0.80 & 20.69 \\
      & k=200, n=200 & 0.49 & 0.80 & 0.49 & 0.81 & 22.02 \\
      & k=350, n=200 & 0.49 & 0.82 & 0.47 & 0.83 & 25.38 \\
      & k=500, n=200 & 0.49 & 0.82 & 0.46 & 0.84 & 27.35 \\
      & k=750, n=500 & 0.50 & 0.81 & 0.45 & 0.85 & 50.58 \\
      & k=1000, n=500 & 0.52 & 0.80 & 0.43 & 0.86 & 57.08 \\
      & sl\_baseline & 0.53 & 0.77 & 0.53 & 0.77 & 1035.43 \\
      & sl\_hdps & 0.48 & 0.82 & 0.47 & 0.83 & 1636.48 \\\hline

      NSAID & k=50, n=200 & 0.60 & 0.68 & 0.61 & 0.67 & 43.15 \\
      & k=100, n=200 & 0.60 & 0.69 & 0.60 & 0.69 & 43.48 \\
      & k=200, n=200 & 0.59 & 0.70 & 0.60 & 0.69 & 47.08 \\
      & k=350, n=200 & 0.60 & 0.69 & 0.59 & 0.70 & 52.99 \\
      & k=500, n=200 & 0.60 & 0.69 & 0.59 & 0.71 & 58.90 \\
      & k=750, n=500 & 0.60 & 0.69 & 0.58 & 0.71 & 112.44 \\
      & k=1000, n=500 & 0.61 & 0.69 & 0.58 & 0.72 & 119.28 \\
      & sl\_baseline & 0.61 & 0.67 & 0.61 & 0.66 & 1101.84 \\
      & sl\_hdps & 0.59 & 0.70 & 0.59 & 0.71 & 2075.05 \\\hline

      VYTORIN & k=50, n=200 & 0.44 & 0.64 & 0.43 & 0.64 & 113.45 \\
      & k=100, n=200 & 0.43 & 0.65 & 0.43 & 0.65 & 116.73 \\
      & k=200, n=200 & 0.43 & 0.65 & 0.43 & 0.66 & 146.81 \\
      & k=350, n=200 & 0.43 & 0.65 & 0.42 & 0.67 & 166.18 \\
      & k=500, n=200 & 0.43 & 0.65 & 0.42 & 0.67 & 189.18 \\
      & k=750, n=500 & 0.43 & 0.65 & 0.42 & 0.68 & 315.22 \\
      & k=1000, n=500 & 0.43 & 0.65 & 0.42 & 0.68 & 350.45 \\
      & sl\_baseline & 0.42 & 0.69 & 0.42 & 0.70 & 9165.93 \\
      & sl\_hdps & 0.42 & 0.70 & 0.41 & 0.71 & 15743.89 \\
      \hline
    \end{tabular}
  }
  \caption{Perfomance for hdPS algorithms and Super Learners}
  \label{res}
\end{table}

\subsection{Tuning Parameters for the hdPS Screening Method}

The screening process of the hdPS needs to be cross-validated in the same step as its predictive algorithm. For this study, the computation is too expensive for this procedure, so there is an additional risk of overfitting due to the selection of hdPS covariates. A solution would be to generate various hdPS covariate sets under different hdPS hyper parameters and fit the machine learning algorithms on each covariate set. Then, SL3 would find the optimal ensemble among all the hdPS covariate set/learning algorithm combinations. 

\subsection{Performance of the hdPS}

Although the hdPS is a simple logistic algorithm, it  takes advantage of extra information from claims data. It is, therefore, reasonable that the hdPS generally outperforms the algorithms that do not take into account this information  in most cases. Processing time for the hdPS is sensitive to $n$ while less sensitive of $k$ (see \ref{table:SL_time}). For the datasets evaluated in this study, however, the hdPS was not sensitive to either $n$ or $k$ (see table \ref{res}). Therefore, Super Learners which include the hdPS may save  processing time by including only a limited selection of hdPS algorithms 
without sacrificing performance.

\subsubsection{Risk of overfitting the hdPS}

\begin{figure}[H]
  \begin{center}
    \includegraphics[width=0.7\textwidth]{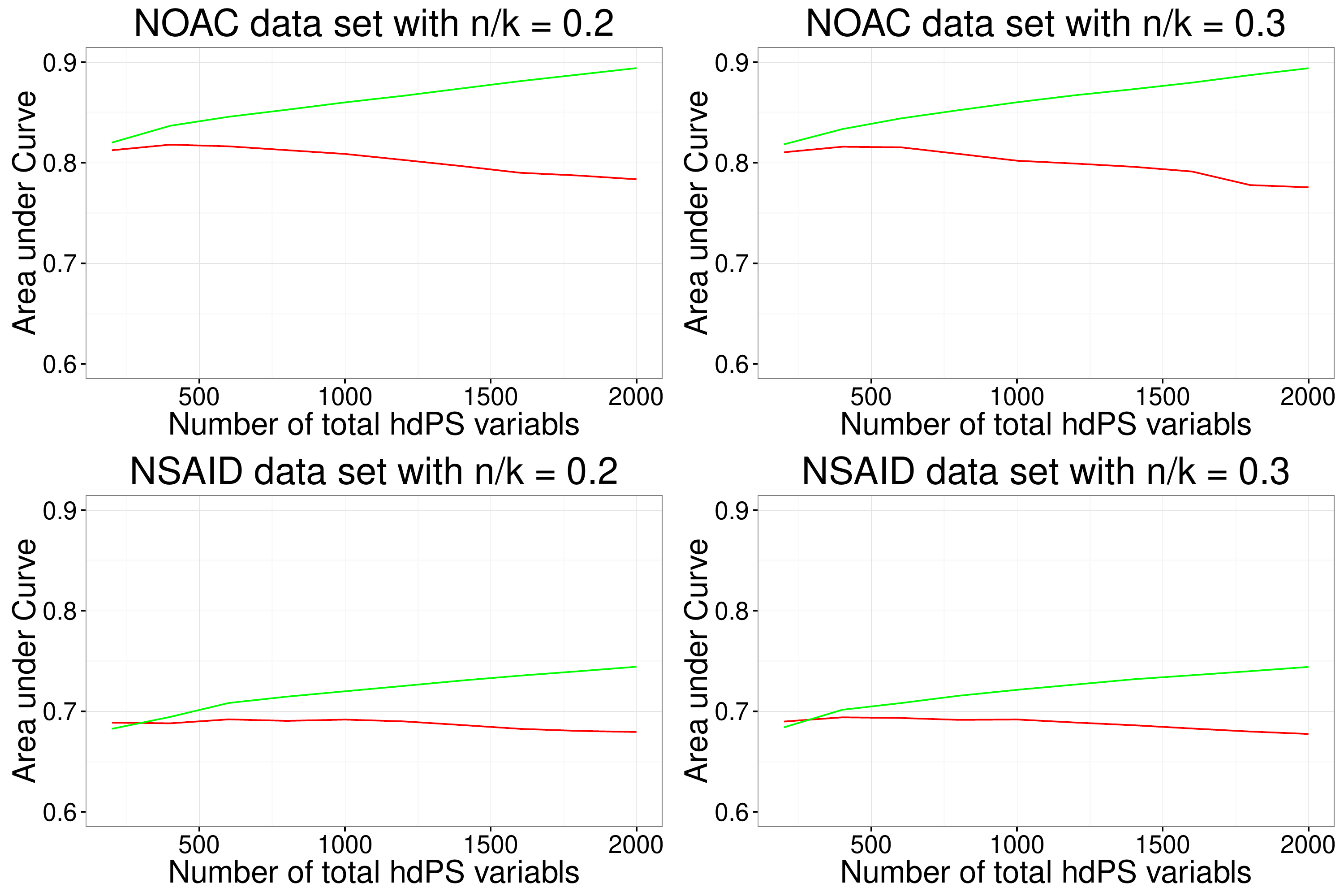}
    \caption{AUC for hdPS algorithms with different number of variables, $k$.}
    \label{overfittingAUC}
  \end{center}
\end{figure}

\begin{figure}[H]
  \begin{center}
    \includegraphics[width=0.7\textwidth]{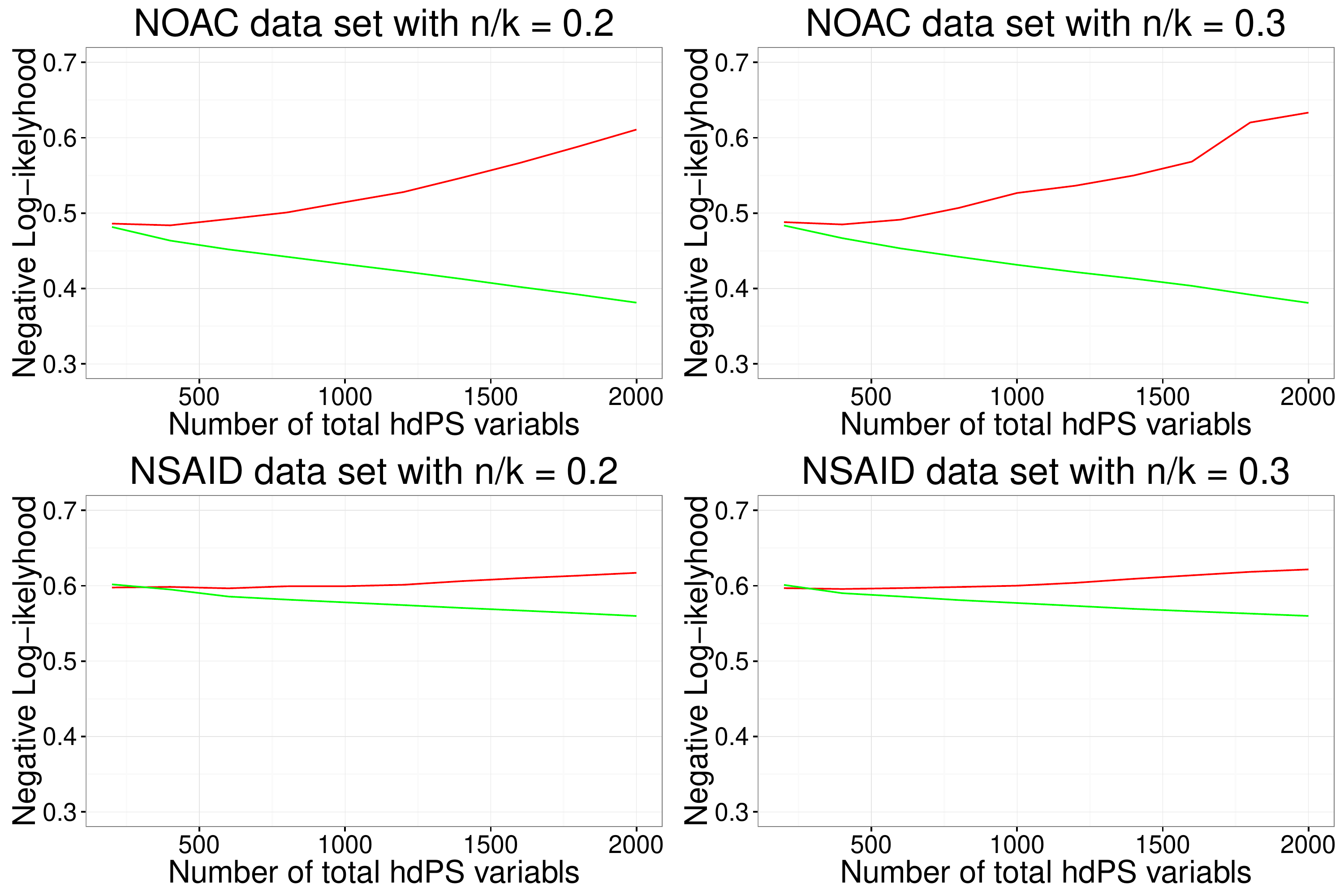}
    \caption{Negative loglikelihood for hdPS algorithms with different number of variables, $k$.}
    \label{overfittingLL}
  \end{center}
\end{figure}

The hdPS algorithm utilizes many more features than traditional
methods, which may raise the risk of overfitting. Table \ref{res} shows the negative loglikelihood for both the training set and  testing set. From Table \ref{res} we see that differences in the performance of the hdPS within the training set and test set are small. This suggests that in these data, performance was not sensitive to small or moderate differences in the specifications for $k$ and $n$. 

To study the impact of overfitting the hdPS across each data set, we fixed the proportion of the number of variables per dimension ($n$) and the number of total hdPS variables ($k$). We then increased $k$ to observe the sensitivity of the performance of the hdPS algorithms. The green lines represent the performance over the training sets and the red lines represent peformance over the test sets.

From figure \ref{overfittingAUC}, we see that increasing the number of variables in the hdPS algorithm results in an increase in AUC in the training sets. This is deterministically a result of 
increasing model complexity. To mitigate this effect, we looked at the AUC over the 
test sets to determine if model complexity reduces performance. For both $n/k = 0.2$ and
$n/k = 0.4$, AUC in the testing sets is fairly stable for $k < 500$, but begins to decrease for larger values of $k$. The hdPS appears to be the most  sensitive to overfitting for $k>500$.

Similarly, in figure \ref{overfittingLL}, the negative log-likelihood decreases  in the training sets as $k$ gets larger, but  begins to increase within the testing sets  for $k > 500$, similar to what we found for AUC. Thus, we conclude that the negative log-likelihood is also less sensitive to $k$ for $k<500$. Therefore, in these datasets the hdPS appears to be sensitive to overfitting only when values of k are greater than 500. Due to the large sample sizes of our datasets, the binary nature of the claims code covariates, and the sparsity of hdPS variables, the hdPS  algorithms are at less of a risk of overfitting. However, the high dimensionality of the data may lead to some computation issues.

\subsubsection{Regularized hdPS}

\begin{figure}[H]
  \begin{center}
    \includegraphics[width=0.8\textwidth]{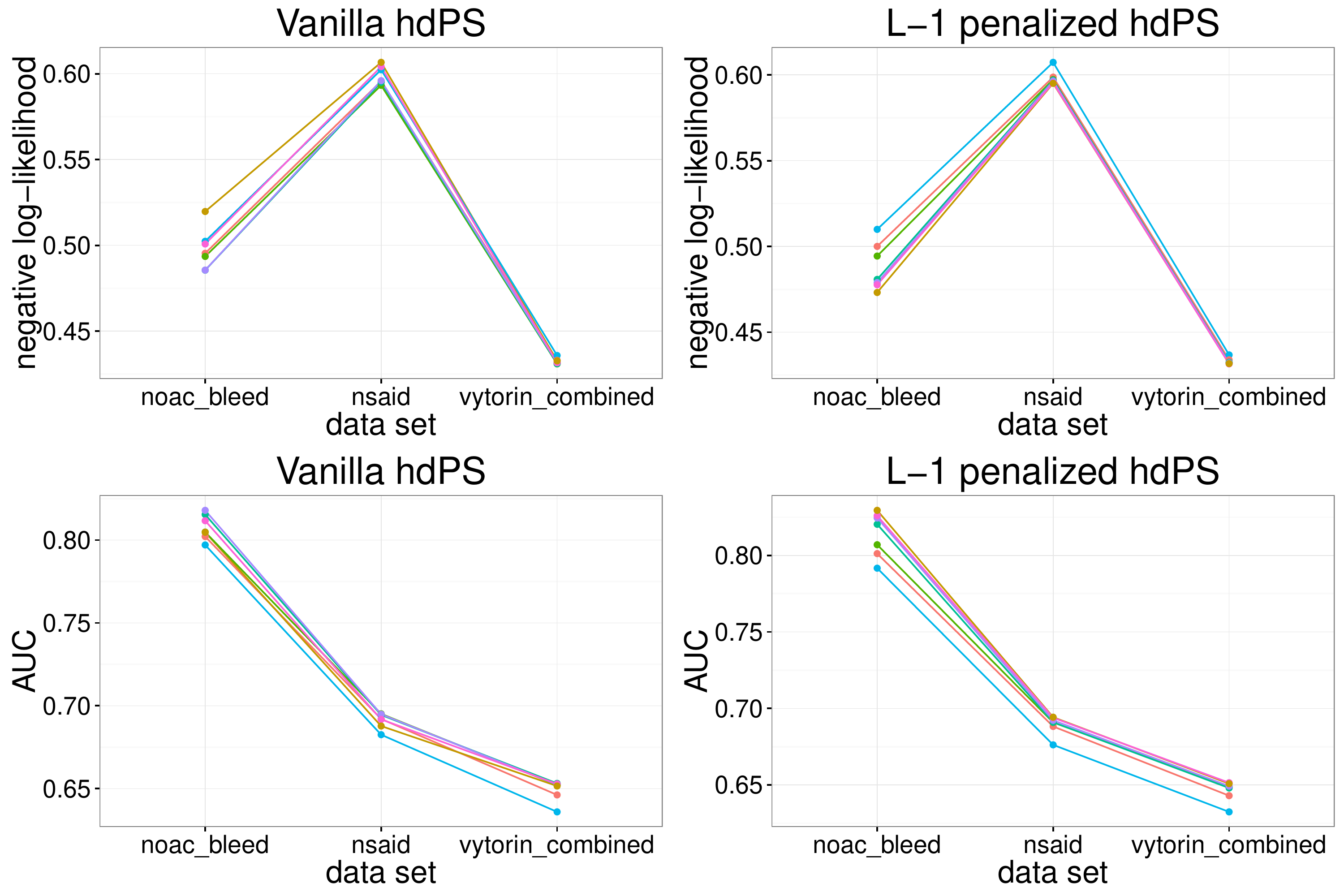}
    \caption{Vanilla (unregularized) hdPS Compared to Regularized hdPS}
    \label{lasso}
  \end{center}
\end{figure}

The hdPS algorithm uses multivariate logistic regression for its
estimation. We compared the performance of this algorithm against
that of regularized regression by implementing the estimation step using the \emph{cv.glmnet} method in  \emph{glmnet} package in R \citep{friedman2009glmnet}, which uses cross-validation to find the best tuning parameter $\lambda$.

To study if regularization can decrease the risk of overfitting the hdPS, we used $L-1$ regularization (LASSO) for the logistic regression step. For every regular hdPS we used cross-validation to find the best tuning parameter based on discrete Super Learner (which selects the model with the tuning parameter that minimizes the cross-validated loss).

Figure \ref{lasso} shows the negative log-likelihood and AUC over the test sets for unregularized hdPS (left) and regularized hdPS (right). We can see that using regularization can increase performance slightly. In this study, the sample size is relatively large  and the benefits of regularization are minimal. However, when dealing with smaller data sets, it is likely that regularized regression will have more of an impact when estimating high-dimensional PSs. Alternatively, one could first generate hdPS covariates and then use Super Learner (as described in SL3).

\subsection{Predictive Performance for SL}

SL is a weighted linear combination of candidate learner algorithms that has been demonstrated to perform asymptotically at least as well as the best choice among the library of candidate algorithms, whether or not the library contains a correctly specified parametric statistical model.  The results from this study are consistent with these theoretical results and demonstrate within large healthcare databases that the SL is optimal in terms of optimizing prediction performance. 

It is interesting that the SL also performed well compared to the individual candidate algorithms in terms of maximizing the AUC. Even though the specified loss function within the SL algorithm was the cross-validated negative log-likelihood, the SL outperformed individual candidate algorithms in terms of the AUC. Finally, for the datasets evaluated in this study, incorporating hdPS generated variables within the SL improved prediction performance. In this study, we found that the hdPS variable selection algorithm provided a simple way to utilize additional information from claims data, which improved the prediction of treatment assignment.

\subsection{Data-adaptive property of SL}

The SL has a number of advantages for the estimation of propensity scores: First, estimating the propensity score using a parametric model requires accepting strong assumptions concerning the functional form of the relationship between treatment allocation and the covariates. Propensity score model misspecification may result in significant bias in the treatment effect estimate \citep{rubin2004principles,brookhart2006variable}. Second, the relative performance of different algorithms relies heavily on the underlying data generating distribution. This paper  demonstrates that no single prediction algorithm is optimal in every setting. Including many different types of algorithms in the SL library accommodates this issue. Cross-validation helps to avoid the risk of overfitting, which can be particularly problematic when modeling high-dimensional sets of variables within small to moderate sized datasets.

To summarize, we found that Gradient Boosting and the hdPS resulted in the dominant weights within the SL algorithm in all three datasets. Therefore, in these examples, these were the two most powerful algorithms for predicting treatment assignment. Future research could explore the performance of only including algorithms with large weights if computation time is limited. Also, this study illustrates that the optimal learner for prediction depends on the underlying data-generating distribution. Including many algorithms within the SL library, including hdPS generated variables, can improve the flexibility and robustness of the SL algorithm when applied to large healthcare databases.

\section{Conclusion}
\label{S:5}

In this study, we thoroughly investigated the performance of the SL for predicting treatment assignment in administrative healthcare databases. Using three empirical datasets, we demonstrated how the SL can adaptively combine information from a number of different algorithms to improve prediction modeling in these settings.

In particular, we introduced a novel strategy that combines the SL with the hdPS variable selection algorithm. We found that the SL can easily take advantage of the extra information provided by the hdPS to improve its flexibility and performance in healthcare claims data. While previous studies have implemented the SL within healthcare claims data, this study is the first to thoroughly investigate its performance in combination with the hdPS within real empirical datasets. We conclude that combining the hdPS with SL prediction modeling is promising for predicting treatment assignment in large healthcare databases.


\bibliographystyle{abbrvnat} 
\bibliography{references}

\appendix

\section*{Appendix}

\begin{tabular}{|p{8cm}|p{3cm}|p{3cm}|}
  \hline
  Model name & Abbreviation & R Package \\ \hline
Bayesian Generalized Linear Model &  bayesglm & arm \\ 
C5.0 & C5.0 & C50, plyr \\ 
Single C5.0 Ruleset & C5.0Rules & C50\\ 
Single C5.0 Tree & C5.0Tree & C50 \\ 
Conditional Inference Tree & ctree2 & party\\  
Multivariate Adaptive Regression Spline & earth & earth\\  
Boosted Generalized Linear Model & glmboost & plyr, mboost\\ 
Penalized Discriminant Analysis & pda & mda\\ 
Shrinkage Discriminant Analysis & sda & sda \\
Flexible Discriminant Analysis & fda & earth, mda\\  
Lasso and Elastic-Net Regularized Generalized Linear Models & glmnet & glmnet\\  
Penalized Discriminant Analysis & pda2 & mda\\  
Stepwise Diagonal Linear Discriminant Analysis & sddaLDA & SDDA\\ 
Stochastic Gradient Boosting  & gbm & gbm, plyr\\ 
Multivariate Adaptive Regression Splines & gcvEarth & earth\\ 
Boosted Logistic Regression & LogitBoost & caTools \\ 
Penalized Multinomial Regression & multinom & nnet\\  
Penalized Logistic Regression & plr & stepPlr\\  
CART & rpart & rpart, plyr,\newline rotationForest\\ 
Stepwise Diagonal Quadratic Discriminant Analysis & sddaQDA & SDDA\\  
Generalized Linear Model & glm & stats \\ 
Nearest Shrunken Centroids & pam & pamr\\  
Cost-Sensitive CART &  rpartCost & rpart\\\hline
\end{tabular}

\end{document}